\documentclass[english,preprint]{revtex4}
\usepackage[T1]{fontenc}
\usepackage[latin1]{inputenc}
\usepackage{graphicx}
\usepackage{amssymb}

\makeatletter

\makeatletter

 \date{\today}

\makeatother

\usepackage{babel}
\makeatother

\begin{document}

\title{Spin dynamics in point contacts to single ferromagnetic films}

\author{O. P. Balkashin$^{1}$, V. V. Fisun$^{1,2}$, I. K. Yanson$^{1}$,
L. Yu. Triputen$^{1}$, A. Konovalenko$^{2}$, and V. Korenivski$^{2}$}

\affiliation{$^{1}$B. Verkin Institute for Low Temperature Physics and Engineering,
National Academy of Sciences of Ukraine, 61103 Kharkiv, Ukraine; }

\affiliation{$^{2}$Nanostructure Physics, Royal Institute of Technology, SE-10691
Stockholm}


\begin{abstract}
Excitation of magnons or spin-waves driven by nominally unpolarized
transport currents in point contacts of normal and ferromagnetic metals
is probed by irradiating the contacts with microwaves. Two characteristic
dynamic effects are observed: a rectification of off-resonance microwave
current by spin-wave nonlinearities in the point contact conductance,
and a resonant stimulation of spin-wave modes in the nano-contact
core by the microwave field. These observations provide a direct evidence
that the magnetoconductance effects observed are due to GHz spin dynamics
at the ferromagnetic interface driven by the spin transfer torque
effect of the transport current.
\end{abstract}
\maketitle
The pioneering predictions of spin transfer torque (STT) effects {[}1,2]
have been confirmed in numerous experiments on ferromagnetic/nonmagnetic
nanostructures. Most of the experiments have been performed on spin-valves
where the current, spin-polarized by a hard ferromagnetic layer of
normalized magnetization $\textbf{m}_{2}=\textbf{M}_{2}/M_{\textrm{s}}$,
produces a spin torque on a magnetically soft layer resulting in a
precession or switching of the soft layer's magnetization ($\textbf{m}_{1}$).
Due to the effect of giant magnetoresistance {[}3] these switching
and precession are translated in to either abrupt hysteretic changes
of the resistance of the tri-layer {[}4,5] or an \emph{ac} voltage
at the frequency of the magnetization precession {[}6-8]. The oscillation
frequency is a function of the magnitude of the driving current -
the effect considered to be highly promising for current controlled
oscillators for use in microelectronics {[}9]. This spin dynamic effect
is often analyzed in the macrospin approximation {[}2,10] using the
Landau-Lifshitz-Gilbert-Slonczewski equation {[}2], in which the additional
torque caused by a spin polarized current of magnitude $I$, counteracting
the intrinsic damping torque, is \begin{equation}
\frac{d\textbf{m}_{1}}{dt}\propto I\eta(\Theta)\textbf{m}_{1}\times[\textbf{m}_{1}\times\textbf{m}_{2}].\label{eq:Torque}\end{equation}
Here $\Theta=\cos^{-1}(\textbf{m}_{1}\cdot\textbf{m}_{2})$ reflects
the degree of the magnetization misalignment for the two layers, and
$\eta(\Theta)$ the effective spin polarization in the system {[}10].
Thus, the current, spin-polarized by $\textbf{m}_{2}$, produces a
torque on $\textbf{m}_{1}$, which can compensate the intrinsic dissipation
in $\textbf{m}_{1}$ and lead to stationary GHz-range oscillations.
Depending on the magnitude of the current, the associated STT can
increase or decrease the amplitude of the oscillation {[}6-8,10].
Micromagnetic simulations based on the above torque mechanism, with
the discretization size of $\sim1$ nm, allow additionally to study
the nonuniform dynamics of magnetization {[}11]. Such micromagnetic
approach is particularly suitable for analyzing magnetic point contacts,
where both the current and the spin distribution can be nonuniform,
with large misalignment angles $\Theta$ within the ferromagnetic
volume adjacent to the contact. 

Several recent experiments observed STT effects in mechanical 10 nm
range needle/surface contacts {[}12] as well as 50 nm range lithographic
contacts {[}5,13] to single continuous ferromagnetic films. Our previous
experiments on such magnetic point contacts {[}14] have demonstrated
the same static magnetoconductance properties as those found in spin-valves,
and confirmed the mechanism of the effect originally proposed in {[}15]
to be the transverse to the electron flow spin-transfer due to impurity
scattering at the magnetic interface in the non-ballistic regime.
The spin misalignment ($\Theta$) in this \emph{single magnetic layer}
case is due to spacial spin perturbations at the interface within
the contact core. In this paper we report an observation of two spin-dynamic
effects in the conductance of magnetic point contacts, which directly
demonstrate the spin precessional nature of the phenomenon. 

The schematic of the experiment is shown in the inset to Fig. 1. %
\begin{figure}
\centering\includegraphics[width=0.7\columnwidth]{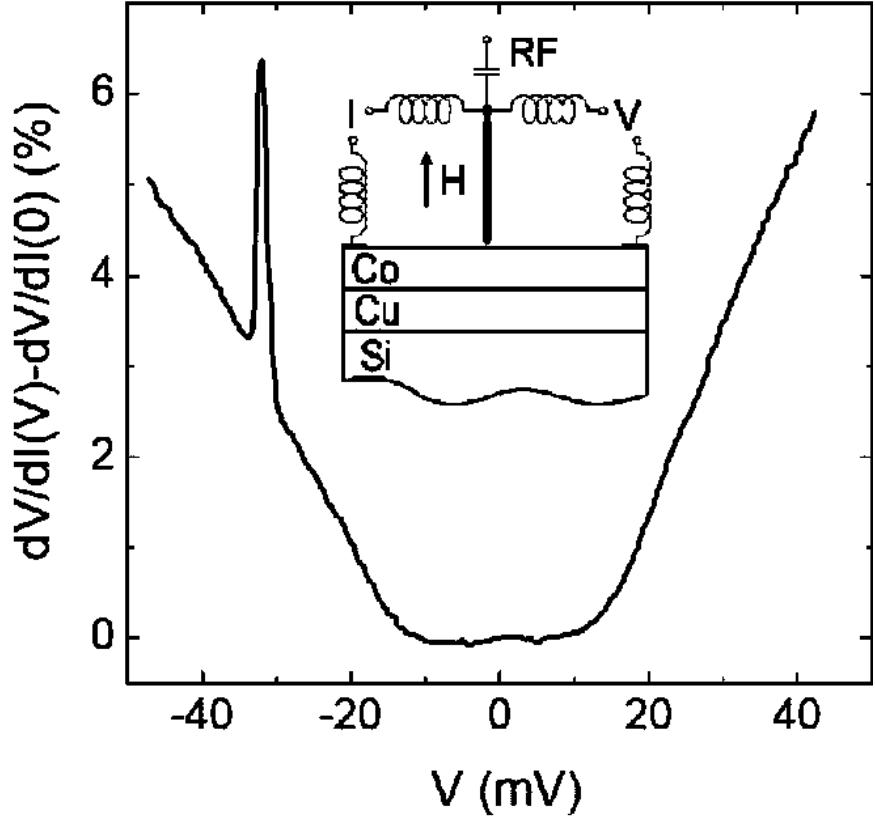}

\caption{Relative differential resistance for point contacts of a Cu needle
and a 100 nm thick Co film, $R_{0}=7.2\ \Omega$, $H_{\perp}=4$ T.
The inset shows the schematic of the experiment.}

\end{figure}
We study point contacts formed nano-mechanically between sharpened
needles of Cu or Ag and Co films of thickness 5, 10, and 100 nm. The
Co films were deposited in ultra high vacuum on oxidized Si substrates
buffered with a 100 nm thick layer of Cu. A subset of the samples
was capped with a 2-3 nm thick layer of Cu or Au to prevent natural
oxidation of Co. Point contacts were formed directly in a liquid He
bath using a two-axis micropositioning mechanism allowing to scan
the surface in selecting the contact location and gradually vary the
mechanical strain on the needle. The contacts had a substantial structural
asymmetry - the needle on the one side and the buffer metal film on
the other. This means that the spin torques acting on the two N$_{1}$/F
and F/N$_{2}$ interfaces did not counter compensate {[}15]. The resistance
of the point contacts varied from 5 to 30 $\Omega$, which using the
standard formula {[}16] corresponds to a diameter range of twenty
to a few nanometers. Differential resistance $dV/dI(V)$ was measured
using a lock-in technique, with the modulating current amplitude of
10-50 $\mu$A and frequency of 443 Hz. The negative \emph{dc} bias
polarity corresponds to the electron current flowing from the needle
into the film. An external magnetic field of up to 5 T was applied
either in or perpendicular to the plane of the films. The microwave
radiation was supplied by directly connecting a coaxial cable to the
point contact electrodes, and decoupling the \emph{dc} and HF circuits
using a bias tee (as shown in the inset to Fig. 1). We have also used
a radiating loop in the form of a short at the end of the coax, which
yielded essentially identical results. This is to be expected for
the PCS technique used, where the needle (top contact) acts as an
efficient antenna at GHz frequencies, converting microwaves into an
\emph{ac} current and vice versa. The maximum RF power in the contact
region was estimated to be a few tens of $\mu$W. The results we report
did not depend on the material of the nonmagnetic needle. In what
follows, we present the data collected using needles of Cu.

The measured differential resistance exhibited peaks characteristic
of the STT effect in normal/ferromagnetic structures. The peaks observed
at negative bias, such as shown in Fig. 1, are not caused by the effect
of the Oersted field of the bias current, which should be symmetric
with respect to the bias polarity. The position of the peak on the
bias axis changed linearly in proportion to the applied field magnitude,
indicating its magnetic origin (see also {[}4,13,14]). Such a sharp
increase in the resistance is highly reproducible for a given contact
and, in the multilayer case, has been conclusively identified as due
to a threshold-like activation of the magnetization precession (spin-wave)
or a change in the magnetization precession angle (a transition into
a different precession mode {[}8]). Some contacts exhibited a more
complex behavior with multiple magneto-conductance maxima and sometimes
even minima, which can be taken as evidence for multiple dynamically
stable magnetization precession modes. Here we report an observation
of two distinct effects in the response of such magneto-conductance
peaks to a microwave radiation of varying frequency and power: a rectification
of off-resonance microwaves by the magnetic nonlinearities in the
conductance, and a resonant stimulation of the peaks by the microwave
field. These observations unambiguously identify the origin of the
phenomenon as a spin precession at the \emph{single} ferromagnetic
interface within the point contact core.

Fig. 2a shows the effect of the RF power of frequency $f=\Omega/2\pi=2$
GHz on the amplitude and shape of a magneto-conductance peak for a
Cu-Co(100 nm) point contact measured in an in-plane magnetic field
of 1.82 T. The magnetic origin of this conductance peak is confirmed
by the field dependence of its positiion shown in the inset, quite
similar to the behavior of the STT peaks in conventional spin-valves.
The initial sharp peak in $dV/dI$ (curve 1) is suppressed and broadened
as the RF power is increased to 24 $\mu$W. The small peak at -7 mV
was diminished in amplitude and smeared out in voltage by microwave
irradiation, which was similar to the behavior of the main peak. The
small amplitude of this additional peak did not allow a quantitative
analysis. We therefore concentrate on the main peak. The RF power
level near the PC was estimated by measuring the transmission coefficient
of the coaxial line. A further increase in the irradiation power leads
to a splitting of the peak. We observed such behavior versus RF power
in many contacts and find it to be highly characteristic of the spin-wave
peaks in $dV/dI$. %
\begin{figure}
\centering\includegraphics[width=0.85\columnwidth]{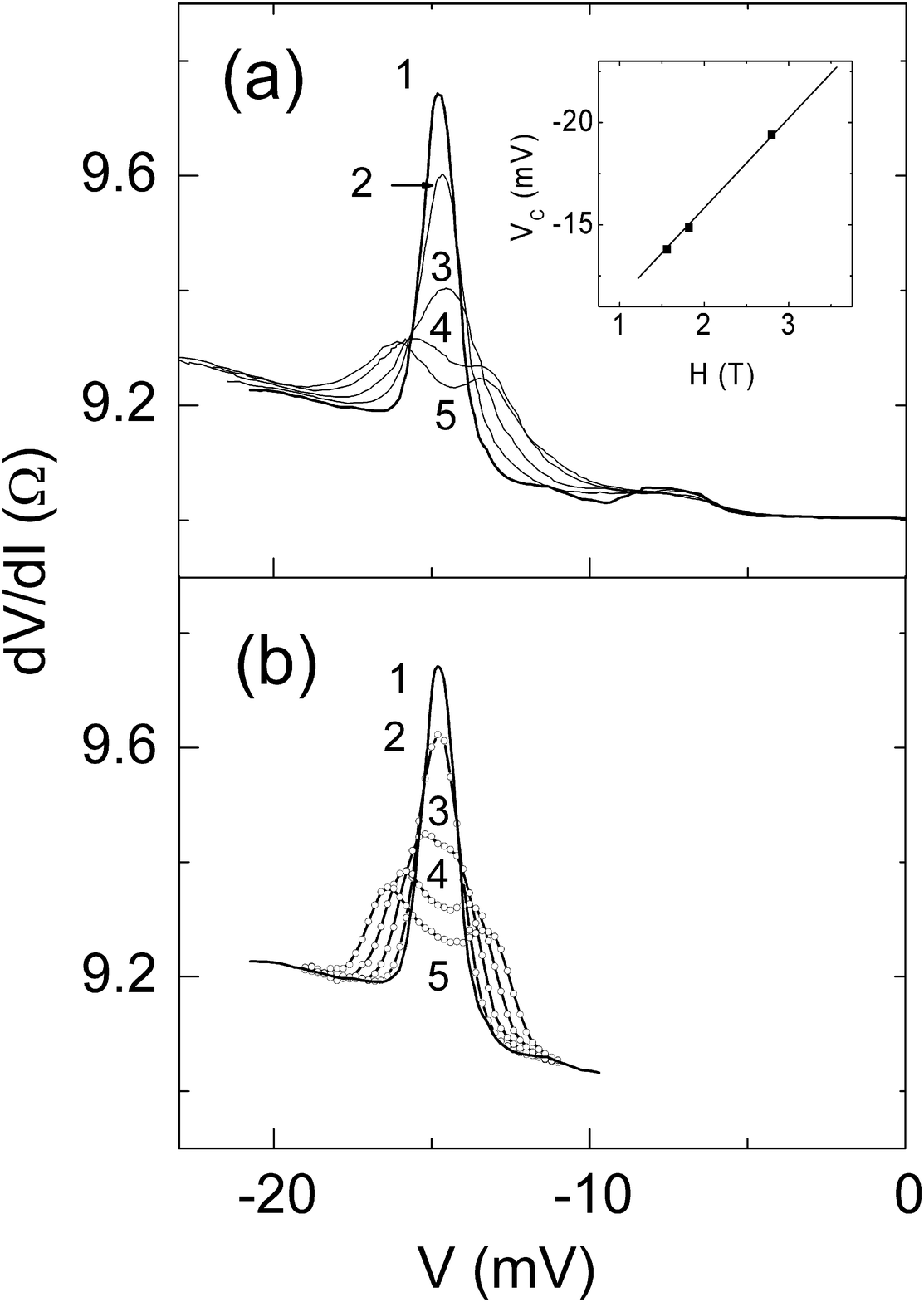}

\caption{(a) $dV/dI$ as a function of bias for a point contact of a Cu needle
and a 100 nm thick Co film under irradiation of frequency 2 GHz and
power $P$=0, 12, 24, 36, and 48 $\mu$W (curves 1, 2, 3, 4, and 5,
respectively). (b) Calculated dependence according to Eq. \ref{eq:I_LF}
for $V_{1}$=0.5, 1, 1.5, and 2 $\mu$V (curves 2, 3, 4, and 5, respectively).
Curves 1 in (a) and (b) are identical. Inset: critical voltage (STT
peak position) as a function of the external field.}

\end{figure}
 To model this behavior we use the the theory of {[}17] for RF irradiated
nonmagnetic point contacts, where the nonlinear conductance is due
to electron-phonon scattering. The results from this work are applicable
to any point contacts where the conductance is due to one electron
processes. In particular, a good agreement with the experiment has
been obtained for superconducting tunnel contacts, semiconductor/metal,
superconductor/normal metal, as well as all metal point contacts {[}18,19].

In the case where a constant bias $V_{0}$ is superposed with a weak
\emph{ac} signal $V_{1}$ of frequency $\Omega$ induced in the contact
by an applied RF power, $V(t)=V_{0}+V_{1}\cos\Omega t$, the time-averaged
current-voltage characteristics (IVC) is given by {[}17,18]

\begin{equation}
\overline{I}(\overline{V})=\frac{\Omega}{\pi}\intop_{0}^{\pi/\Omega}I_{0}(V_{0}+V_{1}\cos\Omega t)dt,\label{eq:I_LF}\end{equation}
where $I_{0}(V_{0})$ is the unperturbed IVC in the absence of RF.
It is assumed that the frequency is low compared to the inverse of
the characteristic electron relaxation time producing the nonlinearity
and that the system is not in resonance. These conditions require
that the energy of the RF photons is much lower than the width of
the nonlinearity in the IVC and that $I_{0}(V_{0})$ is not affected
by the microwaves directly but only through $V_{0}\rightarrow V_{0}+V(t)$
as given by Eq. \ref{eq:I_LF}. In our experiment $\hbar\Omega\sim10^{-2}$
meV, indeed much smaller than the peak half-width in Fig. 2 ($\sim$1
meV). Eq. 2 is the well known result for a classical detector {[}18]
describing \emph{ac} rectification by an \emph{off-resonance} conductance
nonlinearity. The unperturbed $I_{0}(V_{0})$ is obtained by integrating
the experimental $dV/dI(V)$ measured without irradiation (curve 1
in Fig. 2a). Eq. \ref{eq:I_LF} is then used to obtain the IVC expected
under irradiation, with a subsequent numerical differentiation to
obtain the predicted differential resistance. Thus calculated $dV/dI$
for several amplitudes of the \emph{ac} voltage across the contact,
$V_{1}^{2}\sim P$, are shown in Fig. 2b. Curves 1 in Fig. 2a and
2b are identical. An excellent agreement between the measured data
and the predicted behavior is obtained (curves 2-5 in Figs. 2a and
2b), which allows us to conclude that the effect observed is a rectification
of the \emph{off-resonance} microwave current by the magnetic nonlinearity
in the conductance of the nanocontact.

A number of contacts showed a distinctly different behavior. A sharp
peak in the differential resistance appeared under RF irradiation
in an otherwise monotonously increasing dependence, as illustrated
in Fig. 3. This peak is analogous in shape to the one discussed above
(Figs. 1, 2) and was observed only for negative bias polarity. The
effect was completely reversible, with $dV/dI$ returning to its original
monotonous form after the RF irradiation was removed. For some contacts
the original dependence showed a small irregularity in the same region
of bias where later a pronounced peak would develop under irradiation
(curve 1). Other contacts with RF induced peaks showed no signs of
any singularity in the differential resistance in the rf-unperturbed
state. The amplitude of the induced peak increased approximately linearly
with the RF power in the low power range, then saturated at higher
power, as shown in the inset to Fig. 3. Such behavior is expected
for a transition from a low- to a high-angle magnetization precession
with increasing excitation power. It results in a saturation of the
effective precession angle, analogous to the spin response in the
ferromagnetic resonance. Thus, a large angle precession of the spins
localized to the point contact core at the surface of the ferromagnetic
film, with the spins outside the core being essentially static, produces
an additional magnetic contribution to the resistance through the
PC core-film domain-wall magnetoresistance. Our previous experiments
show \cite{nanoletter} that such energetically distinct atomically
thin surface spin layers can form currrent or field driven spin-valves
within a \emph{single} ferromagnetic film.

\begin{figure}
\centering\includegraphics[width=0.75\columnwidth]{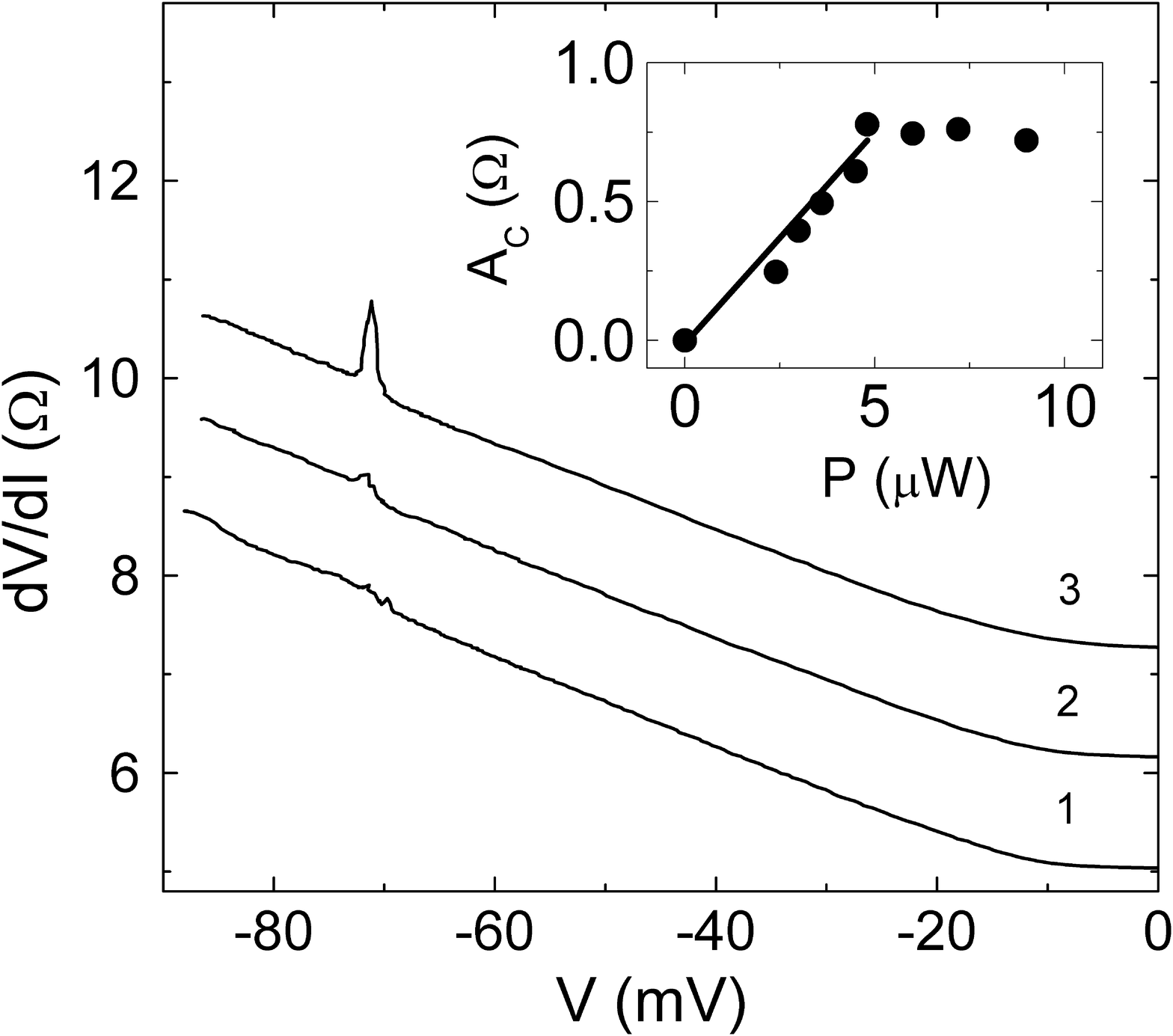}

\caption{Spin-wave peak in $dV/dI$ stimulated by increasing the power of irradiation
at 2 GHz, for RF power $P$=0, 2.4, and 3.6 $\mu$W (curves 1, 2,
and 3, respectively). The curves are shifted vertically by 1 $\Omega$
for clarity. The contact is Cu-Co(100 nm), $R_{0}=5.04\ \Omega$,
$H_{\perp}=2.47$ T. The inset shows the amplitude of the induced
peak as a function of RF power.}

\end{figure}

The critical behavior of an STT-induced magneto-conductance peak under
irradiation is shown in Fig. 4. %
\begin{figure}
\centering\includegraphics[width=0.85\columnwidth]{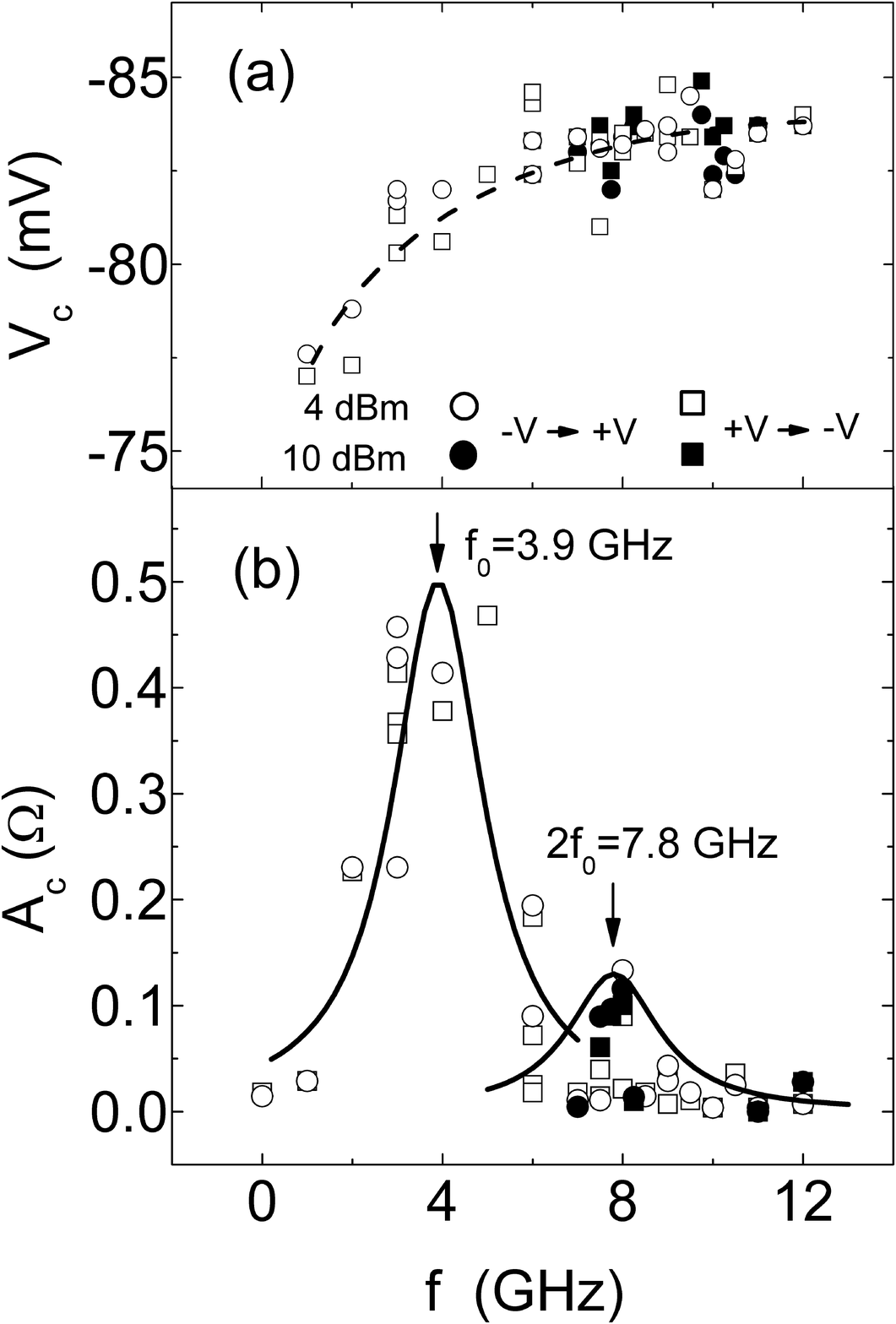}

\caption{Critical voltage (a) and amplitude (b) of a spin-wave peak for a Cu-Co(10
nm) contact with $R_{0}=6.7$ $\Omega$ as a function of the RF frequency
for $P_{rf}=$4 and 10 dBm, open and solid symbols, respectively.
Circles and squares correspond to the positive and negative bias sweep
direction. $H_{\parallel}=1$ T. Lines are Lorenzian peaks approximating
the data, with characteristic frequencies $f_{0}$ and $2f_{0}$.}

\end{figure}
 Repeatedly sweeping the bias and measuring $dV/dI$ for different
radiation frequencies revealed changes in the position and amplitude
of the spin-wave peak, while the monotonic part of the $dV/dI$ was
unchanged. Fig. 4a shows the dependence of the peak position on the
irradiation frequency for two values of the RF power, 4 dBm (1-12
GHz) and 10 dBm (7-12 GHz), measured at the output of the generator.
The higher power allowed to enhance the measured signal, which was
then linearly scaled down to 4 dBm for a direct comparison. The critical
voltage increases with increasing the irradiation frequency (similar
dependence was observed for the externally applied magnetic field).
This observation correlates with the observations and analysis of
the current driven magnetization dynamics in spin-valve structures,
where the critical bias parameters and the spin-torque induced precession
frequency are know to be interdepend {[}7,8]. We observe small changes
in the critical voltage, which are uncorrelated from sweep to sweep,
and are likely caused by small changes in the non-uniform spin states
involved by the strong spin-torques of the transport current. 

The resonant character of the magnetic excitations for single ferromagnetic
layers we observe is most clearly seen in the \emph{nonmonotonic}
dependence of the spin-wave peak amplitude on the irradiation frequency,
shown in Fig. 4b. The observed maxima at 3.9 and 7.8 GHz correspond
to the first and second harmonics of the resonant spin excitation
in the nano-contact core, where the STT is maximum due to the high
current density. These spin excitations likely consist of a whole
spectrum of spin-wave modes since the spin distribution in the contact
is expected to be non-uniform, resulting in broad resonance peaks.
Our resolution is not sufficent for a comparison of the linewidths
of the two harmonics - the ration that can vary widely for highly
non-linear systems \cite{Pufall}. The observed resonance frequencies
of 1-10 GHz discussed above are too low if the resonating object was
a uniformly magnetized film in a field of 1 T. However, the phenomenon
studied in this work differs in a principle way from the uniform film
FMR, and therefore from the majority (if not all) work on spin torques
in multi-layers (see Ref. \cite{berkov} for a recent review, including
spin-valve type point contacts). The spin perturbation (nano-domain,
surface layer, or a spin-vortex) at the core of the point contact,
responsible for the magneto-transport observed, is $\sim$10 nm in
diameter in our case and is significantly different in magnetic properties
from the bulk of the ferromagnetic film. This difference can be due
to a number of factors, such as potentially high mechanical stress
in the point contact core, appreciable magnetostriction (in Co), and
as a result potentially large magnetic anisotropy which can vary significantly
in strength and direction from contact to contact. It is therefore
not surprising that the HF properties are different from those of
the uniform FMR of the underlying ferromagnetic film.

Thus, in addition to the \emph{off-resonance RF rectification} effect,
we report the first observation of a \emph{resonant absorption} of
RF radiation by spin precessional modes in the presence of spin transfer
torques, for \emph{single} ferromagnetic interfaces. The interpretation
of the effect is as follows. When the frequency of the RF field $f$
equals the frequency of the magnetization precession $f_{0}$ - a
function of the magnitude of the bias current through the contact
and the applied magnetic field - a resonant increase in the amplitude
of the precession occurs, corresponding to a transition from predominantly
stochastic oscillations to a stationary precession. Taking into account
the highly nonlinear nature of the system, the resonance condition
is $mf\approx nf_{0}(V_{0},H)$, where $m$ and $n$ are integers.
The energy of the spin sub-system in the contact core increases when
this resonance condition is met, which leads to a new precessional
state with a different trajectory, amplitude and axis of precession
of the spins involved. This causes a sharp change in the domain-wall
magnetoresistance of the contact seen as a peak in $dV/dI$ at the
critical bias $V_{C}$ (or $I_{C}$) corresponding to the resonance
condition. Therefore, the critical bias parameters of a magnetic point
contact in resonance with an RF field are a function of the frequency
of the field.

In this interpretation a higher externally applied magnetic filed
would create a higher effective magnetic field in the point contact,
which would then correspond to a higher characteristic precession
frequency. Thus, magnetic field and RF frequency should are expected
to shift the position of resonant spin-torque peaks in a similar fashion.
This is indeed observed. Figs. 5a and 5b show the differential conductance
for two resonant STT peaks (absent without RF irradiation), where
in one case the peak is shifted to high bias by increasing the RF
frequency and in the other by increasing the external magnetic field.
These field and frequency transitions are fully reversible, and the
peaks vanish when the RF is switched off. %
\begin{figure}
\centering\includegraphics[width=0.7\columnwidth]{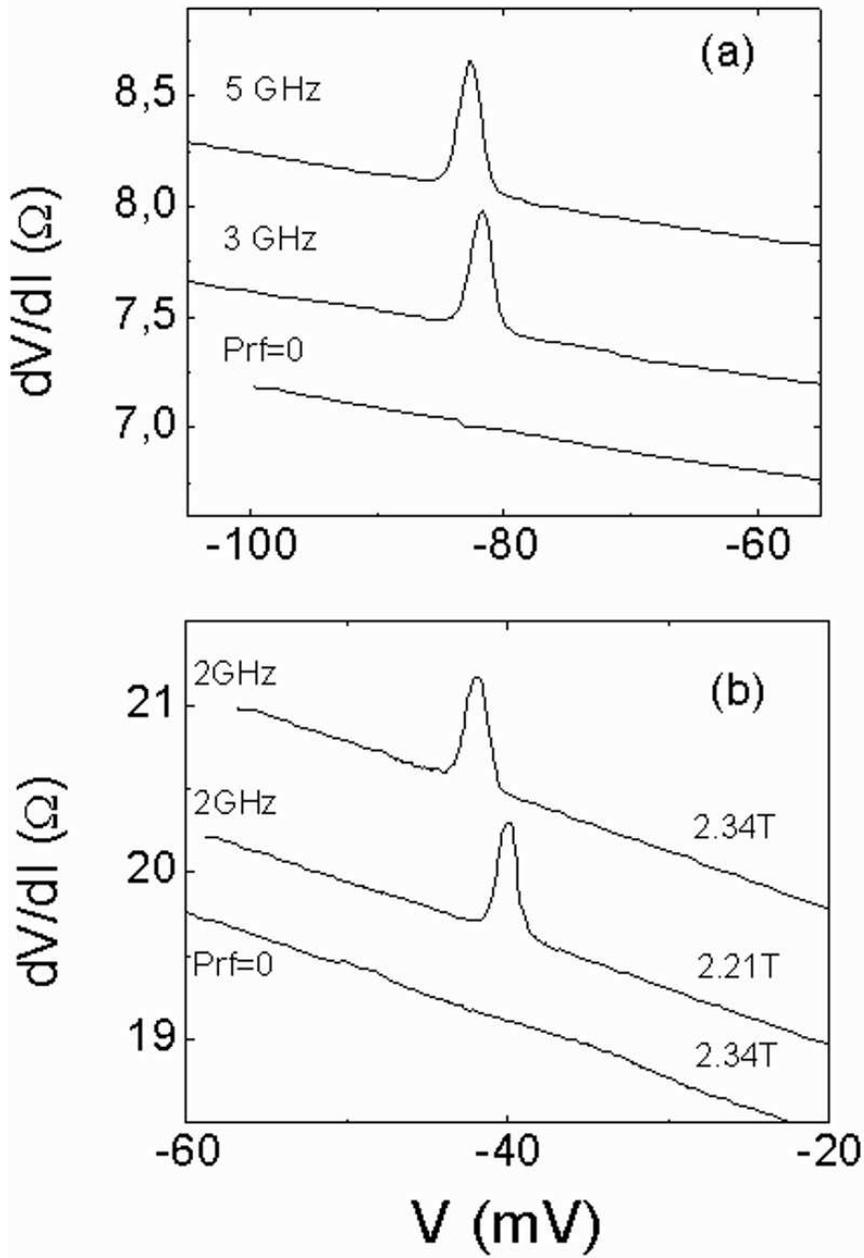}\caption{Differential resistance versus bias voltage for point contacts with
$R$=6.7 $\Omega$ for $f=$3 and 5 GHz (a) and $R$=18 $\Omega$
for $f=$2 GHz $P_{rf}=0$ and 4 dBm (b).}

\end{figure}

Spin-wave excitations in point contacts to Co/Cu multilayers stimulated
by 50 GHz range microwaves were reported in \cite{tsoi}. No off-resonance
\emph{ac} current rectification effects were observed. The shape of
the RF-induced features reported in {[}20] is also significantly different
from the simple resonance like maxima we report. It will be informative
to mention that our tests on point contacts to \emph{multilayer} films
of Co/Cu revealed essentially identical spin torque effects to those
found in \emph{single} layer contacts, showing no dependence on the
thickness of the ferromagnetic films down to a few nm. This strongly
suggests that the effects under consideration are dominated by spin-wave
excitations in the nanocontact core in the immediate vicinity of the
normal/ferromagnetic interface. This is especially true for point
contacts of high resistance, $R\gg1\ \Omega$, much greater that the
'bulk' resistance of the metallic multilayer, where the layered structure
should contribute insignificantly to the measured total resistance.
Our recent spectroscopic study of spin-torque-driven hysteresis in
point contacts to nm-thin ferromagnetic films \cite{nanoletter}
is an additional evidence for the interface rather than bulk nature
of the effects discussed herein.

We would to point out that the technique of point contact spectroscopy
is unique as it allows to study transport in sub-lithographic, often
atomic or near-atomic structures. However, as detailed above, the
point contacts created can vary significantly in they properties.
In this case it is the magnetic anisotropy and the exchange field
profiles that are most relevant. Such a spread in properties can be
a disadvantage in terms of exact \emph{a priori} control of the nano-contact.
In our case of many different nano-contacts analyzed \emph{a posteriori},
the result is an advantage \textendash{} we are able to distinguish
two broad types of behavior, resonant and non-resonant. In one case
the \textquoteleft{}spin dot\textquoteright{} created in the contact
core (with its specific local anisotropy, exchange field profile,
external field strength and direction) can resonantly absorb radiation
of a specific frequency and start a large angle precession which results
in substantial magnetoresistance (appearance of an STT peak). In the
other case, an existing STT peak is suppressed in a predictable way
by an off-resonant irradiation (of sufficient power, non-pecific in
frequency). 

In conclusion, we have probed the high frequency dynamics of the current
induced spin excitations in point contacts to single ferromagnetic
films by irradiating the contacts with microwaves. The observed STT-induced
spin-wave excitations are shown to be of resonant spin precessional
nature. We further show that these spin excitations rectify off-resonance
RF current as theoretically expected. Thus, we demonstrate that the
effect under consideration is the same in its spin-dynamic character
as the STT driven spin dynamics in spin-valves. 

Financial support from the Swedish Foundation for Strategic Research,
Royal Academy of Sciences, and the National Academy of Sciences of
Ukraine under project \char`\"{}Nano\char`\"{} 2/07-H are gratefully
acknowledged.

\end{document}